\documentclass[aps,amsmath,amssymb,preprint,author-year]{revtex4-1}
\usepackage[utf8]{inputenc}
\usepackage{geometry}
\usepackage{mathrsfs}
\usepackage{bm}
\usepackage{amsmath}
\usepackage{graphicx}

\makeatletter

\providecommand{\tabularnewline}{\\}

\@ifundefined{showcaptionsetup}{}{%
 \PassOptionsToPackage{caption=false}{subfig}}
\usepackage{subfig}
\makeatother

\begin{document}
\preprint{KOBE-COSMO-19-02}
\title{Quasinormal modes of $p$-forms in spherical black holes}
\author{Daiske Yoshida}
\email{dice-k.yoshida@stu.kobe-u.ac.jp}

\affiliation{Department of Physics, Kobe University, Kobe, 657-8501, Japan}
\author{Jiro Soda}
\email{jiro@phys.sci.kobe-u.ac.jp}

\affiliation{Department of Physics, Kobe University, Kobe, 657-8501, Japan}
\date{\today}
\begin{abstract}
We study the quasinormal modes of $p$-form fields in spherical black
holes in $D$-dimensions. Using the spherical symmetry of the black
holes and gauge symmetry, we show the $p$-form field can be expressed
in terms of the coexact $p$-form and the coexact $(p-1)$-form on
the sphere $S^{D-2}$. These variables allow us to find the master
equations. By utilizing the S-deformation method, we explicitly show
the stability of $p$-form fields in the spherical black hole spacetime.
Moreover, using the WKB approximation, we calculate the quasinormal
modes of the $p$-form fields in $D(\leq10)$-dimensions. 
\end{abstract}
\pacs{undefined}
\maketitle

\section{Introduction\label{sec:Introduction}}

Black holes in general relativity are important from various perspectives.
In fact, they are sources of gravitational waves, they provide a way
to test general relativity in the strong gravity regime, they can
be a key to quantum gravity. Before discussing these physics, we have
to show the stability of black holes. In fact, the stability of black
holes are often nontrivial. Historically, the stability of black holes
has been studied since the seminal paper by Regge-Wheeler and Zerilli
\citep{Regge:1957td,Zerilli:1970se}, for example, see \citep{Moncrief:1974am,Kay:1987ax,Dafermos:2016uzj}.
From the point of view of a unified theory such as string theory,
it is natural to consider black holes in higher dimensions. Indeed,
higher dimensional black holes may be created at the accelerator such
as the LHC~\citep{Giddings:2001bu}. Thus, the stability analysis
is also generalized to higher dimensions~\citep{Kodama:2003jz,Ishibashi:2003ap,Kodama:2003kk,Kodama:2007ph}.
In higher dimensions, however, Einstein's general relativity is not
a unique possibility. Rather, Lovelock gravity is natural in higher
dimensions~\citep{Lovelock:1971yv,Charmousis:2008kc}. Therefore,
the stability of black holes in Lovelock gravity has been studied~\citep{Takahashi:2009dz,Takahashi:2009xh,Takahashi:2010gz,Takahashi:2010ye,Takahashi:2013thesis,Konoplya:2017lhs}.
Moreover, in contrast to 4-dimensional general relativity where only
scalar, electromagnetic and gravitational field can reside in the
black hole spacetime, there exist $p$-form fields in higher dimensions.
To our best knowledge, no work of $p$-form fields in higher dimensional
black hole spacetime has been done. The purpose of this paper is to
study the stability of $p$-form fields in black hole spacetime and
obtain quasinormal modes of $p$-form fields.

To study the behavior of various physical fields in spherical black
holes, we must derive the master equations. For this purpose, we express
a $p$-form field in terms of a coexact $p$-form and coexact $(p-1)$-form
on the sphere and derive the master equation for each component. If
the effective potential in the master equation is positive outside
event horizon of black holes, the $p$-form field is stable~\citep{Wald:1979nssm}.
However,, it turns out that the effective potential for a $p$-form
field has a negative region for some parameters. This region may cause
the instability of $p$-form fields in spherical black holes. Nevertheless,
we succeed in proving the stability of $p$-form fields using the
S-deformation method~\citep{Ishibashi:2003ap}.

Given the stability, we can calculate the quasinormal modes of $p$-form
fields in the black hole spacetime. We use the WKB method~\citep{Schutz:1985zz,Iyer:1986np,Iyer:1986nq}
to calculate quasinormal modes of $p$-form fields in $D(\leq10)$-dimensions.
Since a $p$-form field has two components, there are two quasinormal
modes, namely, one for each component. It is shown that quasinormal
modes of the $p$-form field in $D$-dimensions reflect duality relations.

The organization of the paper is as follows: In Sec. \ref{sec:p-form},
we review the properties of $p$-form fields. In particular, we count
the physical degrees of freedom of a $p$-form filed. In Sec. \ref{sec:p-form in BH},
we consider $p$-form fields in spherical black holes. We represent
a $p$-form field by coexact form fields on the sphere. We also discuss
spherical harmonics. In Sec. \ref{sec:master}, we obtain the master
equations for $p$-form field in spherical black holes in arbitrary
dimensions. The master variables are a coexact $p$-form and a coexact
$(p-1)$-form on the sphere. We check their degrees of freedom match
to the physical degrees of freedom of a $p$-form field. We also find
useful duality relations for the effective potentials. It turns out
that the effective potential has a negative region for some cases.
In Sec. \ref{sec:Stability-Analysis}, therefore, we have explicitly
verified the stability of the $p$-form field using the $S$-deformation
technique. In Sec. \ref{sec:Quasinormal-Modes}, we also investigated
the quasinormal modes of $p$-form fields using the WKB approximation.
Sec. \ref{sec:Conclusion} is devoted to conclusion.

\section{Basics of $p$-form fields\label{sec:p-form}}

In this section, we introduce the action for $p$-form fields and
explain the gauge invariance in the system. We also calculate physical
degrees of freedom of a $p$-form field. We refer the reader to the
textbook \citep{Choquet:1982book} for more detailed explanation of
the $p$-forms on manifolds.

The action of the $p$-form field is given by 
\begin{align}
S & =-\frac{1}{2}\int\bm{F}\wedge*\bm{F}=-\frac{1}{2\left(p+1\right)!}\int d^{D}x\sqrt{-g}F_{a_{1}\cdots a_{p+1}}F^{a_{1}\cdots a_{p+1}}\ ,
\end{align}
where we used the Hodge operator $*$. Here, $D$ is the dimension
of the spacetime. We defined the $p$-form field $\bm{A}_{p}$ as
follows 
\begin{equation}
\bm{A}_{p}\equiv\frac{1}{p!}A_{a_{1}\cdots a_{p}}dx^{a_{1}}\wedge dx^{a_{2}}\wedge\cdots\wedge dx^{a_{p}}
\end{equation}
and the field strength $\bm{F}$ is defined by 
\begin{equation}
\bm{F}\equiv d\bm{A}.\label{eq:field_strength}
\end{equation}
The operator $d$ is the exterior derivative satisfying the identity
\begin{equation}
d^{2}=0.\label{eq:d2}
\end{equation}
Using the Hodge operator, we can define the coderivative $\delta$
as 
\begin{equation}
\delta\equiv\left(-1\right)^{D(p+1)+1}*d*.
\end{equation}
Note that the coderivative also satisfies the identity 
\begin{equation}
\delta^{2}=0.\label{eq:cd2}
\end{equation}
The equations of motion for the form field can be deduced as 
\begin{equation}
d\bm{F}=0\hspace{1em}{\rm and}\hspace{1em}\delta\bm{F}=0.\label{eq:EOMofForm}
\end{equation}
This system has the symmetry under the dual transformation 
\begin{equation}
\tilde{\bm{F}}=*\bm{F}.
\end{equation}
Because of this symmetry, we do not need to consider $p$-form field
with the rank higher than 
\begin{align}
p_{{\rm max}}=\left[\frac{D}{2}\right]-1=\left[\frac{n}{2}\right].
\end{align}
Here, $\left[\quad\right]$ denotes the Gauss symbol. So, we concentrate
on the form fields $\bm{A}_{p}\left(0\leq p\leq p_{{\rm max}}\right)$
in arbitrary dimensions.

The $p$-form field has the invariance under the gauge transformation
\begin{equation}
\bm{A}_{p}\to\tilde{\bm{A}}_{p}=\bm{A}_{p}+d\bm{\Xi}_{p-1}.
\end{equation}
with an arbitrary $(p-1)$-form field $\bm{\Xi}_{p-1}$. This is because
the field strength is defined by \eqref{eq:field_strength}. The gauge
parameter $\bm{\Xi}_{p-1}$ itself has the degeneracy 
\begin{equation}
\bm{\Xi}_{p-1}\to\tilde{\bm{\Xi}}_{p-1}=\bm{\Xi}_{p-1}+d\bm{\Xi}_{p-2}.
\end{equation}
with an arbitrary $(p-2)$-form field $\bm{\Xi}_{p-2}$. Hence, in
order to count the physical degrees of freedom of the $p$-form field
$\bm{A}_{p}$, we need to take into account these degrees generated
by the transformations $\bm{\Xi}_{p-1},\bm{\Xi}_{p-2},\cdots\bm{\Xi}_{0}$.
Taking into account that components $A_{0a_{2}\cdots a_{p}}$ are
not dynamical, the formula for physical degrees of freedom is given
by 
\begin{eqnarray}
_{D-1}C_{p}-{}_{D-1}C_{p-1}+{}_{D-1}C_{p-2}-\cdots={}_{D-2}C_{p}\ .
\end{eqnarray}
For example, in $D=4$, a 2-form field has one physical degree of
freedom.

\section{$p$-form fields in spherical black holes \label{sec:p-form in BH}}

In this subsection, we study the decomposition of the $p$-form field
in black hole spacetime in terms of form fields on the sphere. Then,
we eliminate some components using gauge transformations. We also
discuss eigenvalues of spherical harmonics for $p$-forms.

In general relativity, the spherical black hole is known as the Schwarzschild
black hole. It is not difficult to generalize the Schwarzschild black
hole to higher dimensions $D>4$, and the solutions are called Schwarzschild-Tangherlini
black hole \citep{Tangherlini1963} expressed by the metric 
\begin{equation}
ds^{2}=-f\left(r\right)dt^{2}+\frac{1}{f\left(r\right)}dr^{2}+r^{2}q_{AB}dx^{A}dx^{B}\ ,
\end{equation}
where $f\left(r\right)$ is given by 
\begin{equation}
f\left(r\right)=1-\frac{\mu}{r^{n-1}},
\end{equation}
and $n$ is defined as $n\equiv D-2$. Here, $q_{AB}$ is the metric
of the sphere $S^{n}$ and the spherical coordinate is expressed by
\begin{equation}
x^{A}=\left(\theta^{1},\cdots,\theta^{n}\right).
\end{equation}

\subsection{Decomposition of a $p$-form field in terms of coexact form fields
on the sphere}

We use the notation $\hat{\bm{A}}_{p}$ to denote a $p$-form field
on the sphere, that is, $\hat{\bm{A}}_{p}$ is written by 
\begin{align}
\hat{\bm{A}}_{p} & =A_{A_{1}\cdots A_{p}}\mathscr{D}^{A_{1}\cdots A_{p}}\ ,
\end{align}
where 
\begin{equation}
\mathscr{D}^{A_{1}\cdots A_{p}}\equiv dx^{A_{1}}\wedge\cdots\wedge dx^{A_{p}}.
\end{equation}
We can write down the components of $\bm{A}_{p}$ as follows, 
\begin{align}
\bm{A}_{p}= & \frac{1}{\left(p-2\right)!}A_{trA_{1}\cdots A_{p-2}}dt\wedge dr\wedge\mathscr{D}^{A_{1}\cdots A_{p-2}}+\frac{1}{\left(p-1\right)!}A_{tA_{1}\cdots A_{p-1}}dt\wedge\mathscr{D}^{A_{1}\cdots A_{p-1}}\nonumber \\
 & +\frac{1}{\left(p-1\right)!}A_{rA_{1}\cdots A_{p-1}}dr\wedge\mathscr{D}^{A_{1}\cdots A_{p-1}}+\frac{1}{p!}A_{A_{1}\cdots A_{p}}\mathscr{D}^{A_{1}\cdots A_{p}}.
\end{align}
If we define the components of $\bm{A}_{p}$ as 
\begin{align}
U_{A_{1}\cdots A_{p-2}} & \equiv A_{trA_{1}\cdots A_{p-2}},\\
V_{A_{1}\cdots A_{p-1}} & \equiv A_{tA_{1}\cdots A_{p-1}},\\
W_{A_{1}\cdots A_{p-1}} & \equiv A_{rA_{1}\cdots A_{p-1}},\\
X_{A_{1}\cdots A_{p}} & \equiv A_{A_{1}\cdots A_{p}},
\end{align}
then $\bm{A}_{p}$ is written by 
\begin{equation}
\bm{A}_{p}=dt\wedge dr\wedge\hat{\bm{U}}_{p-2}+dt\wedge\hat{\bm{V}}_{p-1}+dr\wedge\hat{\bm{W}}_{p-1}+\hat{\bm{X}}_{p}.
\end{equation}
Thus, the $p$-form can be expressed by the one $(p-2)$-form $\hat{\bm{U}}_{p-2}$,
the two $(p-1)$-forms $\hat{\bm{V}}_{p-1}$ and $\hat{\bm{W}}_{p-1}$
and the one $p$-form $\hat{\bm{X}}_{p}$. The identity 
\begin{eqnarray}
_{D}C_{p}={}_{D-2}C_{p-2}+2{}_{D-2}C_{p-1}+{}_{D-2}C_{p}
\end{eqnarray}
guarantees the matching of degrees of of freedom.

Using the Hodge decomposition on the sphere $S^{n}$, a $p$-form
field $\hat{\bm{A}}_{p}$ can be decomposed as 
\begin{equation}
\hat{\bm{A}}_{p}=\hat{d}\hat{\bm{A}}_{p-1}+\hat{{\cal A}}_{p},\hspace{1em}\hspace{1em}\text{for}\hspace{1em}1\leq p<n-1\ ,
\end{equation}
where we have introduced coexact form $\hat{\delta}\hat{{\cal A}}_{p}=0$.
From this decomposition theorem on the sphere, we can write down more
useful expansion. In fact, for $p\geq2$, we can express $\hat{\bm{A}}_{p}$
by coexact form, 
\begin{align}
\hat{\bm{A}}_{p} & =\hat{d}\hat{\bm{A}}_{p-1}+\hat{{\cal A}}_{p}\nonumber \\
 & =\hat{d}\left(\hat{d}\hat{\bm{A}}_{p-2}+\hat{{\cal A}}_{p-1}\right)+\hat{{\cal A}}_{p}\nonumber \\
 & =\hat{d}\hat{{\cal A}}_{p-1}+\hat{{\cal A}}_{p}.\label{eq:DecomPform}
\end{align}
This result shows that the general form field on $S^{n}$ is expressed
by only the coexact form fields. Thus, the arbitrary $p$-form field
$\bm{A}_{p}$ is given by 
\begin{align}
\bm{A}_{p}= & dt\wedge dr\wedge\hat{\bm{U}}_{p-2}+dt\wedge\hat{\bm{V}}_{p-1}+dr\wedge\hat{\bm{W}}_{p-1}+\hat{\bm{X}}_{p}\nonumber \\
= & dt\wedge dr\wedge\left(\hat{d}\hat{{\cal U}}_{p-3}+\hat{{\cal U}}_{p-2}\right)+dt\wedge\left(\hat{d}\hat{{\cal V}}_{p-2}+\hat{{\cal V}}_{p-1}\right)\nonumber \\
 & +dr\wedge\left(\hat{d}\hat{{\cal W}}_{p-2}+\hat{{\cal W}}_{p-1}\right)+\left(\hat{d}\hat{{\cal X}}_{p-1}+\hat{{\cal X}}_{p}\right).
\end{align}

\subsection{Gauge fixing of $p$-form field $\bm{A}_{p}$}

The $p$-form field $\bm{A}_{p}$ have the gauge invariance under
the transformation by $\bm{B}_{p-1}$, that is, 
\begin{equation}
\bm{A}_{p}\to\tilde{\bm{A}}_{p}=\bm{A}_{p}+d\bm{B}_{p-1}.
\end{equation}
Now starting from the general expression for $\bm{A}_{p}$ in $D$
dimensions, we can deduce the following result 
\begin{align}
\bm{A}_{p}= & dt\wedge dr\wedge\hat{{\cal U}}_{p-2}+dt\wedge\hat{{\cal V}}_{p-1}+dr\wedge\hat{{\cal W}}_{p-1}+\hat{{\cal X}}_{p}\label{eq:DecompositionOfForm}
\end{align}
by using the gauge transformation for $\bm{A}_{p}$.

\subsection{Spherical harmonics for coexact $p$-form field}

We review the spherical harmonics of the $p$-form field following
\citep{CAMPORESI199457}. Laplace-Beltrami operator is defined by
\begin{equation}
\hat{\Delta}\equiv\hat{\delta}\hat{d}+\hat{d}\hat{\delta}.
\end{equation}
The spherical harmonics of the coexact $p$-form field $\hat{\bm{\mathcal{Y}}}_{p}$
is defined by 
\begin{equation}
\hat{\bm{\mathcal{Y}}}_{p}=\frac{1}{p!}\mathcal{Y}_{A_{1}\cdots A_{p}}\mathscr{D}^{A_{1}\cdots A_{p}}
\end{equation}
which satisfies 
\begin{equation}
\hat{\delta}\hat{\bm{\mathcal{Y}}}_{p}=0
\end{equation}
and 
\begin{equation}
\hat{\Delta}\hat{\bm{\mathcal{Y}}}_{p}=\lambda_{p}\hat{\bm{\mathcal{Y}}}_{p}.
\end{equation}
The last equation becomes 
\begin{equation}
\hat{\delta}\hat{d}\hat{\bm{\mathcal{Y}}}_{p}=\lambda_{p}\hat{\bm{\mathcal{Y}}}_{p}
\end{equation}
by using the identity \eqref{eq:cd2}. Here, $\lambda_{p}$ is given
by 
\begin{equation}
\lambda_{p}\equiv\left(\ell+p\right)\left(\ell+n-p-1\right),
\end{equation}
and $\ell$ is a positive integer, $\ell=1,2,\cdots,\infty$. On the
sphere $S^{n}$, the left hand side of equation becomes 
\begin{align}
\hat{\delta}\hat{d}\hat{\bm{\mathcal{Y}}}_{p} & =\frac{1}{p!}\left(-\left(\tilde{\Delta}-p\left(n-p\right)\right)\mathcal{Y}_{A_{1}\cdots A_{p}}\right)\mathscr{D}^{A_{1}\cdots A_{p}}.
\end{align}
Here, we defined $\tilde{\Delta}$ as 
\begin{equation}
\tilde{\Delta}\mathcal{Y}_{A_{1}\cdots A_{p}}=\mathcal{Y}_{A_{1}\cdots A_{p}}{}^{:A}{}_{:A}.
\end{equation}
Then, the spectrum of the coefficient of the $p$-form harmonics $\hat{\bm{\mathcal{Y}}}_{p}$
is 
\begin{equation}
-\left(\tilde{\Delta}-p\left(n-p\right)\right)\mathcal{Y}_{A_{1}\cdots A_{p}}=\lambda_{p}\mathcal{Y}_{A_{1}\cdots A_{p}}.
\end{equation}
We rewrite this equation as follows 
\begin{equation}
\tilde{\Delta}\mathcal{Y}_{A_{1}\cdots A_{p}}=-\gamma_{p}^{\left(n\right)}\mathcal{Y}_{A_{1}\cdots A_{p}}\ ,
\end{equation}
where 
\begin{equation}
\gamma_{p}^{\left(n\right)}\equiv\lambda_{p}-p\left(n-p\right).
\end{equation}
With the harmonics $\hat{\bm{\mathcal{Y}}}_{p}$, the coefficient
of the general coexact $p$-form field $\hat{{\cal A}}_{p}$ can be
expanded as 
\begin{equation}
{\cal A}_{A_{1}\cdots A_{p}}\left(x\right)=\sum_{l,\sigma}{\cal A}_{\ell,\sigma}\left(t,r\right)\mathcal{Y}_{A_{1}\cdots A_{p}}^{\ell,\sigma}\left(x^{A}\right)\ ,
\end{equation}
where $\sigma$ denotes other indices to characterize the degeneracy.
For simplicity, we denote ${\cal A}_{\ell,\sigma}\left(t,r\right)$
just as ${\cal A}$.

\section{Master equations for $p$-form field \label{sec:master}}

In this section, we derive the master equations for the master variable
$\Psi$ in the Schr$\ddot{{\rm o}}$dinger form 
\begin{equation}
-\ddot{\Psi}+\partial_{x}^{2}\Psi-V\Psi=0\ ,
\end{equation}
where $V$ is the effective potential and $x$ is the tortoise coordinate
defined by 
\begin{equation}
\frac{d}{dx}=f\frac{d}{dr}.
\end{equation}

In eq.\eqref{eq:EOMofForm}, the first equation is trivially satisfied
from the identity \eqref{eq:d2}, and the second equation $\delta\bm{F}=0$
in the coordinate basis is expressed as follows 
\begin{equation}
\frac{1}{\sqrt{-g}}\partial_{a}\left(\sqrt{-g}F^{aa_{1}\cdots a_{p}}\right)=0\ .
\end{equation}
This equation can be decomposed into four patterns.

The first pattern we consider is 
\begin{equation}
\frac{1}{\sqrt{-g}}\partial_{a}\left(\sqrt{-g}F^{atrA_{1}\cdots A_{p-2}}\right)=0\ .
\end{equation}
Substituting the components into the above equation, we obtain 
\begin{equation}
-\frac{1}{r^{2\left(p-1\right)}}\left(\tilde{\Delta}-\left(p-2\right)\left(n-\left(p-2\right)\right)\right){\cal U}^{A_{1}\cdots A_{p-2}}=0
\end{equation}
This yields 
\begin{equation}
\left(\ell+p-2\right)\left(\ell+1+n-p\right){\cal U}_{\ell,\sigma}=0\ .
\end{equation}
Since we are considering $p\geq2$, and generally $n>p$, the quantity
$\left(\ell+p-2\right)\left(\ell+1+n-p\right)$ is always positive.
Hence, the coefficient ${\cal U}_{\ell,\sigma}$ must vanish for all
$\ell$, namely, 
\begin{equation}
\hat{{\cal U}}_{p-2}=0.\label{eq:EoM-p2}
\end{equation}
Thus, the $(p-2)$-form $\hat{{\cal U}}_{p-2}$ in eq.\eqref{eq:DecompositionOfForm}
is not dynamical.

The second pattern is the following: 
\begin{equation}
\frac{1}{\sqrt{-g}}\partial_{a}\left(\sqrt{-g}F^{atA_{1}\cdots A_{p-1}}\right)=0\ .
\end{equation}
It is easy to get 
\begin{equation}
\frac{1}{r^{n-2\left(p-1\right)}}f\partial_{r}\left(r^{n-2\left(p-1\right)}\left(\dot{{\cal W}}-{\cal V}'\right)\right)+\frac{1}{r^{2}}\left(\gamma_{p-1}^{\left(n\right)}+\left(p-1\right)\left(n-p+1\right)\right){\cal V}=0\ .\label{eq:EoM-(p-1)-1}
\end{equation}

The third pattern is given by 
\begin{equation}
\frac{1}{\sqrt{-g}}\partial_{a}\left(\sqrt{-g}F^{arA_{1}\cdots A_{p-1}}\right)=0.
\end{equation}
This leads to 
\begin{equation}
\ddot{{\cal W}}-\dot{{\cal V}}'+\frac{f}{r^{2}}\left(\gamma_{p-1}^{\left(n\right)}+\left(p-1\right)\left(n-p+1\right)\right){\cal W}=0\ .\label{eq:EoM-(p-1)-2}
\end{equation}

The final pattern 
\begin{equation}
\frac{1}{\sqrt{-g}}\partial_{a}\left(\sqrt{-g}F^{aA_{1}\cdots A_{p}}\right)=0
\end{equation}
generates two equations for the $p-1$-form component ${\cal V}^{A_{1}\cdots A_{p-1}}$,
${\cal W}^{A_{1}\cdots A_{p-1}}$ and $p$-form component ${\cal X}^{A_{1}\cdots A_{p}}$
as 
\begin{equation}
\dot{{\cal V}}-\frac{1}{r^{n-2p}}f\partial_{r}\left(r^{n-2p}f{\cal W}\right)=0\label{eq:EoM-(p-1)-3}
\end{equation}
and 
\begin{equation}
-\ddot{{\cal X}}+\frac{1}{r^{n-2p}}f\partial_{r}\left(r^{n-2p}f{\cal X}'\right)-\frac{f}{r^{2}}\left(\gamma_{p}^{\left(n\right)}+p\left(n-p\right)\right){\cal X}=0.\label{eq:EoM-p}
\end{equation}

\subsection{Master equations}

Now, we can derive the master equations. These general results reproduce
the effective potential in eq. (15) and (16) in \citep{Cardoso:2003vt}
and in eq. (99) in \citep{Chaverra:2012bh} as special cases and agree
with the results in \citep{Du:2004jt,LopezOrtega:2006my}.

\subsubsection{Coexact $p$-form}

From eq.\eqref{eq:EoM-p}, we can read off the master variable for
the $p$-form component as 
\begin{equation}
\Psi_{p}=\frac{1}{r^{a}}{\cal X}\hspace{1em}\text{and}\hspace{1em}a\equiv\frac{2p-n}{2}
\end{equation}
Assuming the time dependence of $\Psi_{p}$ as 
\[
e^{-i\omega t},
\]
we obtain the master equation 
\begin{equation}
-\partial_{x}^{2}\Psi_{p}+V_{p,p}^{\left(n\right)}\Psi_{p}=\omega^{2}\Psi_{p}
\end{equation}
with the effective potential $V_{p,p}^{\left(n\right)}$ 
\begin{align}
V_{p,p}^{\left(n\right)}=\frac{f}{r^{2}}\left(\left(\ell+p\right)\left(\ell+n-p-1\right)+\frac{n-2p}{2}\left(rf'+\frac{n-2p-2}{2}f\right)\right).
\end{align}

\subsubsection{Coexact $(p-1)$-form}

The $(p-1)$-form components is contained in the eq.\eqref{eq:EoM-(p-1)-1},
\eqref{eq:EoM-(p-1)-2} and \eqref{eq:EoM-(p-1)-3}, but the master
equation is derived by the eq.\eqref{eq:EoM-(p-1)-2} and \eqref{eq:EoM-(p-1)-3}.
Using the master variable $\Psi_{p-1}$ for the $(p-1)$-form 
\begin{equation}
\Psi_{p-1}\equiv\frac{f}{r^{a}}{\cal W}\hspace{1em}\text{and}\hspace{1em}{\cal V}=r^{2a}f\partial_{r}\left(\frac{f}{r^{2a}}\int{\cal W}dt\right).
\end{equation}
we can deduce the master equation as 
\begin{equation}
-\partial_{x}^{2}\Psi_{p-1}+V_{p,p-1}^{\left(n\right)}\Psi_{p-1}=\omega^{2}\Psi_{p-1}.
\end{equation}
Here, we assumed the same time dependence as before. The effective
potential $V_{p,p-1}^{\left(n\right)}$ is given by 
\begin{equation}
V_{p,p-1}^{\left(n\right)}=\frac{f}{r^{2}}\left(\left(\ell+p-1\right)\left(\ell+n-p\right)+\frac{n-2p}{2}\left(\frac{n-2p+2}{2}f-rf'\right)\right).
\end{equation}
Note that Eq.\eqref{eq:EoM-(p-1)-1} is trivially satisfied.

\subsection{Degrees of freedom and dual relations}

We have shown that $D$-dimensional $p$-form field can be represented
by coexact $p$-form and coexact $(p-1)$-form fields. The condition
for coexact form $\delta\hat{{\cal A}}_{p}=0$ can be solved as 
\begin{equation}
\hat{{\cal A}}_{p}=\delta\hat{{\cal B}}_{p+1}\ .
\end{equation}
However, $\hat{{\cal B}}_{p+1}$ has a freedom $\hat{{\cal B}}_{p+1}+\delta\hat{{\cal B}}_{p+2}$.
This argument continues up to the maximum value $n$. Hence, the degrees
of freedom of the coexact $p$-form is given by 
\begin{equation}
_{n}C_{p+1}-{}_{n}C_{p+2}+{}_{n}C_{p+3}+\cdots={}_{n}C_{p}-{}_{n}C_{p-1}+{}_{n}C_{p-2}+\cdots={}_{n-1}C_{p}\ .
\end{equation}
Similarly, we obtain $_{n-1}C_{p-1}$ for the degrees of freedom of
coexact $(p-1)$-form. Note that the identity 
\begin{equation}
_{n-1}C_{p}+{}_{n-1}C_{p-1}={}_{n}C_{p}
\end{equation}
exactly coincides with the physical degrees of freedom of a $p$-form
field.

The duality plays an important role in form fields. Indeed, we found
the following duality relations 
\begin{eqnarray}
 & V_{n-p,n-p}^{\left(n\right)}=V_{p,p-1}^{\left(n\right)},\label{duality_relation1}\\
 & V_{n-p,n-p-1}^{\left(n\right)}=V_{p+1,p+1}^{\left(n\right)}\ .\label{duality_relation2}
\end{eqnarray}
In particular, in even dimensions, we have the degeneracy 
\begin{eqnarray}
V_{\frac{n}{2},\frac{n}{2}}^{\left(n\right)}=V_{\frac{n}{2},\frac{n}{2}-1}^{\left(n\right)}\ .
\end{eqnarray}
Later, we will see this degeneracy in the quasinormal mode spectrum.

\subsection{Examples of effective potentials}

In $4$ dimensions where $n=2$, the master equation for the $p=0$-form
$\bm{A}_{0}$ becomes 
\begin{align}
V_{0,0}^{\left(2\right)} & =\frac{f}{r^{2}}\left(\ell\left(\ell+1\right)+rf'\right).\label{eq:EffV-p0_0in4}
\end{align}
From the master equations of the $p=1$-form $\bm{A}_{1}$, we see
the effective potential of the coexact $1$-form is 
\begin{align}
V_{1,1}^{\left(2\right)} & =\ell\left(\ell+1\right)\frac{f}{r^{2}}
\end{align}
and that of coexact 0-form reads 
\begin{align}
V_{1,0}^{\left(2\right)} & =\ell\left(\ell+1\right)\frac{f}{r^{2}}.
\end{align}
The effective potentials for the the coexact $1$-form and the coexact
$0$-form components in $\bm{A}_{1}$ are same. Our results agree
with the expression (99) in \citep{Chaverra:2012bh}. Moreover, since
the coexact $2$-forms on the sphere $S^{2}$ does not exist, the
master equation for the $p=2$-form $\bm{A}_{2}$ becomes single.
The effective potential of the coexact 1-form component is given by
\begin{align}
V_{2,1}^{\left(2\right)} & =\frac{f}{r^{2}}\left(\ell\left(\ell+1\right)+rf'\right)\ .
\end{align}
As is expected, this effective potential is the same expression as
that for the $0$-form field in eq.\eqref{eq:EffV-p0_0in4}. This
confirms we can consider only the form fields with the rank larger
than $p_{M}$.

In 5 dimensions where $n=3$, the master equation for the $p=0$-form
$\bm{A}_{0}$ becomes 
\begin{align}
V_{0,0}^{\left(3\right)} & =\frac{f}{r^{2}}\left(\ell\left(\ell+2\right)+\frac{3}{2}\left(rf'+\frac{f}{2}\right)\right).\label{eq:EffV-p0_0in5}
\end{align}
We can also read off the effective potentials in the master equations
for the $1$-form $\bm{A}_{1}$. The effective potential of the coexact
1-form component reads 
\begin{align}
V_{1,1}^{\left(3\right)} & =\frac{f}{r^{2}}\left(\left(\ell+1\right)^{2}+\frac{1}{2}\left(rf'-\frac{1}{2}f\right)\right)\label{eq:EffV-p1_1in5}
\end{align}
and that of the coexact 0-form component is given by 
\begin{align}
V_{1,0}^{\left(3\right)} & =\frac{f}{r^{2}}\left(\ell\left(\ell+2\right)+\frac{1}{2}\left(\frac{3}{2}f-rf'\right)\right).\label{eq:EffV-p1_0in5}
\end{align}
In 5-dimensions, we need not consider a 2-form field because of the
duality. Here, for the check, we look at the master equation for the
2-form $\bm{A}_{2}$. The effective potentials of the coexact 2-form
is given by 
\begin{align}
V_{2,2}^{\left(3\right)} & =\frac{f}{r^{2}}\left(\ell\left(\ell+2\right)+\frac{1}{2}\left(\frac{3}{2}f-rf'\right)\right)\label{eq:EffV-p2_2in5}
\end{align}
and that of the coexact 1-form component becomes 
\begin{align}
V_{2,1}^{\left(3\right)} & =\frac{f}{r^{2}}\left(\left(\ell+1\right)^{2}+\frac{1}{2}\left(rf'-\frac{1}{2}f\right)\right).\label{eq:EffV-p2_1in5}
\end{align}
The effective potential \eqref{eq:EffV-p2_1in5} is the same as eq.\eqref{eq:EffV-p1_1in5},
and the effective potential \eqref{eq:EffV-p2_2in5} is the same as
eq.\eqref{eq:EffV-p1_0in5}. These results just reflect the duality
relations \eqref{duality_relation1} and \eqref{duality_relation2}.

It is also easy to explicitly write down the master equations for
the $p$-form fields in higher dimensions.

\section{Stability Analysis\label{sec:Stability-Analysis}}

\begin{figure}
\begin{centering}
\includegraphics{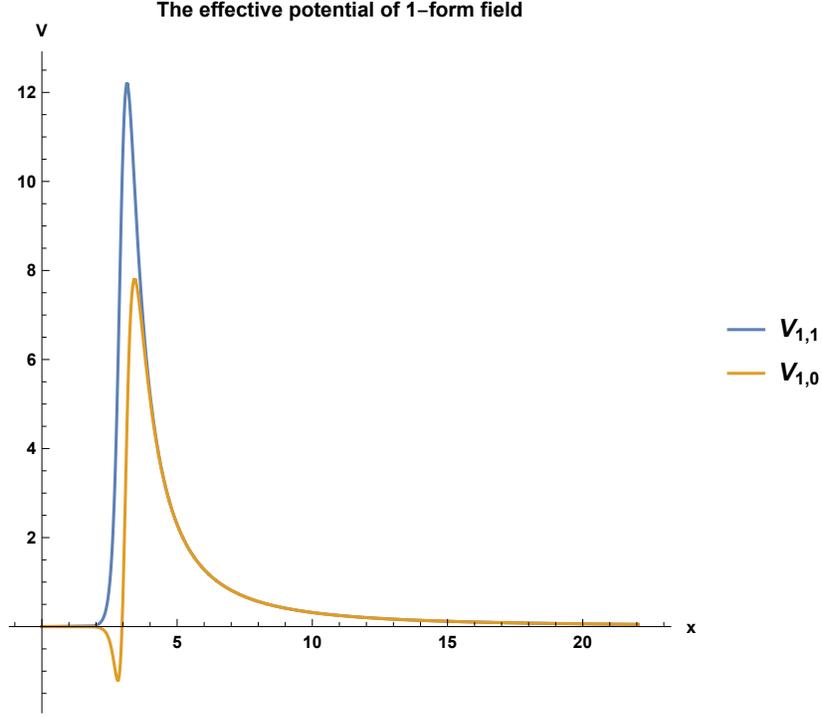} 
\par\end{centering}
\caption{The effective potentials for $1$-form in $10$-dimensions are plotted.
The potential for the coexact $0$-form component has a negative region.
Here, we took $\ell=1$ and $\mu=1$.}
\label{figure_1} 
\end{figure}

In this section, we show the stability of the $p$-form field in arbitrary
dimensions. The stability of $p$-form fields in black hole spacetime
is non-trivial because the effective potential has a negative region
as is shown in Fig.\ref{figure_1}. To show the stability of the fields
around the black holes, the S-deformation method \citep{Kodama:2003jz,Ishibashi:2003ap,Kodama:2003kk,Kodama:2007ph,Kimura:2017uor}
is useful. Hence, first, we shortly review the S-deformation method.
Secondly, we show the stability of the effective potential for each
component of the $p$-form field. 

Let us start with the master equation 
\begin{align}
\omega^{2}\Psi & =A\Psi\ ,\label{eq:operated_master_eq}
\end{align}
where we defined the operator $A$ is 
\begin{equation}
A\equiv-\frac{d^{2}}{dx^{2}}\Psi+V\Psi,
\end{equation}
and we assume the time-dependence of $\Psi$ as 
\begin{equation}
\Psi\propto e^{-i\omega t}.
\end{equation}
If $\omega^{2}<0$ for the boundary conditions $\Psi\to0$ and $d\Psi/dx\to0$
at $x\to\pm\infty$, this solution is unstable and exponentially grows.
So, if we want to show the stability, we have to show $\omega^{2}>0$.
The S-deformation is defined by using the new operator $D_{x}$ as
follows 
\begin{equation}
D_{x}\equiv\frac{d}{dx}+S\left(x\right).
\end{equation}
Then, eq. \eqref{eq:operated_master_eq} is modified as 
\begin{align}
\omega^{2}\left|\Psi\right|^{2} & =\Psi^{*}A\Psi\nonumber \\
 & =\Psi^{*}\left[-\frac{d^{2}}{dx^{2}}+V\right]\Psi\nonumber \\
 & =-\frac{d}{dx}\left(\Psi^{*}D_{x}\Psi\right)+\left|D_{x}\Psi\right|^{2}+\bar{V}\left|\Psi\right|^{2},\hspace{1em}\bar{V}\equiv V+\frac{dS}{dx}-S^{2}.
\end{align}
Suppose we find the continuous function, $S$, which makes $\bar{V}>0$
for $-\infty<x<\infty$. Then, after integration, we have 
\begin{equation}
\omega^{2}\int_{-\infty}^{\infty}dx\left|\Psi\right|^{2}=-\left[\Psi^{*}D_{x}\Psi\right]_{-\infty}^{\infty}+\int_{-\infty}^{\infty}dx\left[\left|D_{x}\Psi\right|^{2}+\bar{V}\left|\Psi\right|^{2}\right]>0\label{eq:inequality}
\end{equation}
Here, we assumed the boundary conditions $\Psi\to0$ and $d\Psi/dx\to0$
at $x\to\pm\infty$, that is the $\Psi$ and $d\Psi/dx$ have a compact
support, so the first term in eq.\eqref{eq:inequality} vanishes.
The above inequality implies the positivity of $\omega^{2}$, that
is, there is no growing mode for $\Psi$.

In the present case, an appropriate S-deformation exists as follows
\begin{equation}
S=-\frac{d}{dx}\left(\ln\left(\frac{1}{r^{\alpha}}\right)\right)
\end{equation}
which is inspired by \citep{Takahashi:2013thesis}.

\subsubsection{Effective potential of coexact $p$-form }

If we choose 
\begin{equation}
\alpha=-\frac{n-2p}{2},
\end{equation}
the effective potential of the coexact $p$-form on the sphere becomes
\begin{align}
V_{p,p}^{\left(n\right)}\to\bar{V}_{p,p}^{\left(n\right)}= & V_{p,p}^{\left(n\right)}+f\frac{dS}{dr}-S^{2}\nonumber \\
= & \frac{f}{r^{2}}\left(\ell+p\right)\left(\ell+n-p-1\right).
\end{align}
This modified potential is positive definite outside the event horizon,
because $p$ satisfies $p\leq p_{{\rm max}}<n$.

\subsubsection{Effective potential of coexact $(p-1)$-form}

If we choose 
\begin{equation}
\alpha=\frac{n-2p}{2},
\end{equation}
the effective potential of the coexact $(p-1)$-form on the sphere
becomes 
\begin{align}
V_{p,p-1}^{\left(n\right)}\to\bar{V}_{p,p-1}^{\left(n\right)}= & V_{p,p-1}^{\left(n\right)}+f\frac{dS}{dr}-S^{2}\nonumber \\
= & \frac{f}{r^{2}}\left(\left(\ell+p\right)\left(\ell+n-p-1\right)-\left(n-2p\right)\right)\nonumber \\
\geq & \frac{f}{r^{2}}\left(\left(1+p\right)\left(1+n-p-1\right)-\left(n-2p\right)\right)\nonumber \\
= & \frac{f}{r^{2}}p\left(n-p+1\right)>0.
\end{align}
This modified potential is also positive definite outside the event
horizon.

From the above analysis, we see the $p$-form fields in arbitrary
dimensions are stable in the spherical black hole even for other gravity
theories as long as $f$ satisfies the positivity $f>0$ outside the
horizon $r>r_{{\rm h}}$.

\section{Quasinormal Modes\label{sec:Quasinormal-Modes}}

We got the master equations and the effective potentials for each
component of the $p$-form field. In this section we present the quasinormal
modes for the general form fields. The quasinormal modes are fundamental
vibration modes around a black hole, and these modes are obtained
by solving the master equation under the appropriate boundary conditions.
The general formalism for calculating quasinormal modes by using WKB
approximation has been proposed by Schutz and Will in \citep{Schutz:1985zz}
and subsequently developed by many people in~\citep{Iyer:1986np,Iyer:1986nq,Kokkotas:1988fm,Seidel:1989bp,Konoplya:2003ii,Konoplya:2011qq,Yoshida:2015vua}.
Here, we summarize the main points of the WKB method for calculating
quasinormal modes.

Firstly, we can divide the region into two regions. The region I ($-\infty<x<x_{0}$)
is the one ranging from top of the effective potential $x_{0}$ to
the horizon of the black hole. The region II ($x_{0}<x<\infty$) is
the one ranging from top of the potential $x_{0}$ to the the far
outside of the black hole, i.e., infinity. The wave traveling to the
potential is called in-going wave and the wave traveling from the
potential is called out-going wave. In each region, the solutions
of the master equation can be expressed as 
\begin{equation}
\begin{cases}
\Psi_{{\rm I}}=Z_{{\rm I}}^{{\rm in}}\Psi_{{\rm I}}^{{\rm in}}+Z_{{\rm I}}^{{\rm out}}\Psi_{{\rm I}}^{{\rm out}}\ ,\\
\Psi_{{\rm II}}=Z_{{\rm II}}^{{\rm in}}\Psi_{{\rm II}}^{{\rm in}}+Z_{{\rm II}}^{{\rm out}}\Psi_{{\rm II}}^{{\rm out}}\ ,
\end{cases}
\end{equation}
where $\Psi^{{\rm in}}$ and $\Psi^{{\rm out}}$ represent ingoing
and outgoing waves, respectively. The boundary condition for obtaining
quasi-normal modes is that there is no in-going waves: 
\begin{equation}
Z_{{\rm I}}^{{\rm in}}=Z_{{\rm II}}^{{\rm in}}=0.
\end{equation}
Since there are two conditions, only discrete complex eigenvalues
are allowed. 
\begin{widetext}
In the $N$th order WKB-method, we approximate the function $Q\left(x\right)$
defined by 
\begin{equation}
Q\left(x\right)\equiv\omega^{2}-V\left(x\right)
\end{equation}
in terms of $2N$th order Taylor expansion around the maximum of the
potential as follows; 
\begin{equation}
\left.\frac{\mathrm{d}Q\left(x\right)}{\mathrm{d}x}\right|_{x=x_{0}}=Q_{0}^{\left(1\right)}=0\hspace{5mm}{\rm and}\hspace{5mm}Q\left(x\right)\simeq\sum_{k=0}^{2N}\frac{1}{k!}Q_{0}^{\left(k\right)}\left(x-x_{0}\right)^{k}.
\end{equation}
Expressing the wave function using WKB approximation, we can calculate
scattering matrix. Thus, we can get the formula for quasinormal modes
as  
\begin{equation}
\omega\simeq\sqrt{V_{0}+\sqrt{2V_{0}^{(2)}}\left(n_{{\rm tone}}+\frac{1}{2}+\sum_{k=1}^{N-1}\Omega_{k}\right)}\ ,
\end{equation}
where $V_{0}\equiv V\left(x_{0}\right)$, and $V_{0}^{(k)}$ is the
$k$th order derivative of the potential 
\begin{equation}
V_{0}^{(k)}\equiv\left.\frac{d^{k}V\left(x\right)}{dx^{k}}\right|_{x=x_{0}},
\end{equation}
and the first and the second of $\Omega_{k}$ are given by  
\begin{align}
\Omega_{1} & =\left(-30\,\beta_{1}^{2}+6\beta_{2}\right)\beta^{2}-\frac{7}{2}\beta_{1}^{2}+\frac{3}{2}\beta_{2}\\
\nonumber \\
\Omega_{2} & =\left(-2820\beta_{1}^{4}+1800\beta_{1}^{2}\beta_{2}-280\beta_{1}\beta_{3}-68\beta_{2}^{2}+20\beta_{4}\right)\beta^{3}\nonumber \\
 & \hspace{1em}+\left(-1155\beta_{1}^{4}+918\beta_{1}^{2}\beta_{2}-190\beta_{1}\beta_{3}-67\beta_{2}^{2}+25\beta_{4}\right)\beta
\end{align}
Here, $\beta$ is defined by 
\begin{equation}
\beta\equiv n_{{\rm tone}}+\frac{1}{2}
\end{equation}
and $\beta_{k}\left(k\geq1\right)$ is defined by 
\begin{equation}
\beta_{k}\equiv\frac{V_{0}^{\left(k+2\right)}}{\left(k+2\right)!}\left(\frac{1}{2V_{0}^{\left(2\right)}}\right)^{\frac{k}{4}+1}.
\end{equation}
The higher $\Omega_{k}$ can be derived explicitly, but the expression
of $\Omega_{k}\left(k\geq3\right)$ is too long to write down here.
So, we write them in Appendix. The parameter, $n_{{\rm tone}}$, is
called tone number of quasi-normal modes. This method is often called
the $N$th order WKB approximation. It is known that in the case of
$n_{{\rm tone}}<\ell$ this approximation is good. So we focus on
the case $n_{{\rm tone}}=0$ in this paper. 
\end{widetext}

\begin{table}[h]
\subfloat[The $2$-form component]{\begin{centering}
\begin{tabular}{|c|c|}
\hline 
$\hspace{1em}D\hspace{1em}$  & QNM: $\omega$\tabularnewline
\hline 
\hline 
$6$  & $\hspace{1em}1.2618-0.4616i\hspace{1em}$\tabularnewline
\hline 
$7$  & $\hspace{1em}1.7509-0.5920i\hspace{1em}$\tabularnewline
\hline 
$8$  & $\hspace{1em}2.2231-0.7192i\hspace{1em}$\tabularnewline
\hline 
$9$  & $\hspace{1em}2.6681-0.8555i\hspace{1em}$\tabularnewline
\hline 
$10$  & $\hspace{1em}3.0791-1.0080i\hspace{1em}$\tabularnewline
\hline 
\end{tabular}
\par\end{centering}
}\subfloat[The 1-form component]{\begin{centering}
\begin{tabular}{|c|c|}
\hline 
$\hspace{1em}D\hspace{1em}$  & QNM: $\omega$\tabularnewline
\hline 
\hline 
$6$  & $\hspace{1em}1.2618-0.4616i\hspace{1em}$\tabularnewline
\hline 
$7$  & $\hspace{1em}1.5387-0.5652i\hspace{1em}$\tabularnewline
\hline 
$8$  & $\hspace{1em}1.8352-0.7345i\hspace{1em}$\tabularnewline
\hline 
$9$  & $\hspace{1em}2.2630-0.8521i\hspace{1em}$\tabularnewline
\hline 
$10$  & $\hspace{1em}2.6910-0.9444i\hspace{1em}$\tabularnewline
\hline 
\end{tabular}
\par\end{centering}
}

\caption{The QNMs of $2$-form field $\bm{A}_{2}$}
\label{table-1} 
\end{table}

\begin{table}[h]
\subfloat[The $3$-form component]{\begin{centering}
\begin{tabular}{|c|c|}
\hline 
$\hspace{1em}D\hspace{1em}$  & QNM: $\omega$\tabularnewline
\hline 
\hline 
$8$  & $\hspace{1em}2.0779-0.6754i\hspace{1em}$\tabularnewline
\hline 
$9$  & $\hspace{1em}2.6018-0.7640i\hspace{1em}$\tabularnewline
\hline 
$10$  & $\hspace{1em}3.1539-0.8307i\hspace{1em}$\tabularnewline
\hline 
\end{tabular}
\par\end{centering}
}\subfloat[The 2-form component]{\begin{centering}
\begin{tabular}{|c|c|}
\hline 
$\hspace{1em}D\hspace{1em}$  & QNM: $\omega$\tabularnewline
\hline 
\hline 
$8$  & $\hspace{1em}2.0779-0.6754i\hspace{1em}$\tabularnewline
\hline 
$9$  & $\hspace{1em}2.3795-0.7729i\hspace{1em}$\tabularnewline
\hline 
$10$  & $\hspace{1em}2.6947-0.9197\hspace{1em}$\tabularnewline
\hline 
\end{tabular}
\par\end{centering}
}

\caption{The QNMs of 3-form field $\bm{A}_{3}$}
\label{table-1-1} 
\end{table}

\begin{table}[h]
\subfloat[The $4$-form component]{\begin{centering}
\begin{tabular}{|c|c|}
\hline 
$\hspace{1em}D\hspace{1em}$  & QNM: $\omega$\tabularnewline
\hline 
\hline 
$10$  & $\hspace{1em}2.9227-0.8595i\hspace{1em}$\tabularnewline
\hline 
\end{tabular}
\par\end{centering}
}\subfloat[The 3-form component]{\begin{centering}
\begin{tabular}{|c|c|}
\hline 
$\hspace{1em}D\hspace{1em}$  & QNM: $\omega$\tabularnewline
\hline 
\hline 
$10$  & $\hspace{1em}2.9227-0.8595i\hspace{1em}$\tabularnewline
\hline 
\end{tabular}
\par\end{centering}
}

\caption{The QNMs of $4$-form field $\bm{A}_{4}$}
\label{table-1-2} 
\end{table}

Now, we show the quasinormal modes of the $p$-form fields up to $D=10$-dimensions.
In this case, we need to consider form fields up to $p=4$. In this
study we used the 6th order WKB method, so we need the $\Omega_{1},\Omega_{2},\cdots,\Omega_{5}$.
We choose a set of parameters, 
\begin{equation}
\left(\ell,\mu,n_{{\rm tone}}\right)=\left(1,1,0\right).
\end{equation}

In Table \ref{table-1}, we showed the quasinormal modes (QNMs) of
a $2$-form field. As you can see QNM of coexact 2-form component
and QNM of coexact 1-form component in 6-dimensions coincide. This
comes from the duality relations \eqref{duality_relation1} and \eqref{duality_relation2}.
Except for $D=8$, the coexact 2-form component decays faster than
coexact 1-form component. In Table \ref{table-1-1}, we listed QNMs
of a $3$-form field. In $D=8$ dimensions, we can see duality relations
hold. In other dimensions, the coexact 2-form component decays faster
than coexact 3-form component. In Table \ref{table-1-2}, we displayed
QNMs of a $4$-form field. In this case, only $D=10$ is relevant.
Here, we see the duality relations again. In all cases, we see, as
$D$ increases, real and imaginary parts of quasinormal frequency
increase.

\section{Conclusion\label{sec:Conclusion}}

We studied the quasinormal modes of $p$-form fields in spherical
black holes in arbitrary dimensions. Using the spherical symmetry
of the black holes and gauge symmetry, we showed the $p$-form field
can be expressed in terms of the coexact $p$-form and coexact $(p-1)$-form
on the sphere. These variables allow us to find the master equations.
We revealed some relations between the effective potentials in the
master equations. We found the effective potential can have a negative
region for some parameters. Therefore, by utilizing the S-deformation
method, we explicitly showed the stability of $p$-form fields in
the spherical black hole spacetime. Finally, using the WKB approximation,
we calculated the quasinormal modes of $p$-form fields in $D(\leq10)$-dimensions.
There, we can see the degeneracy of the spectrum expected from the
duality relations we found.

It is interesting to include rotations of black holes in our analysis.
Recently, Lunin found the ansatz for the $1$-form field in arbitrary
dimensions to get the separable equations of motion~\citep{Lunin:2017drx}.
It is interesting to investigate $p$-form fields in higher dimensional
rotational black holes. We can also consider higher spin fields in
arbitrary dimensions. Resolving the above problems must have implications
for string theory. We leave these problems for future work. 
\begin{acknowledgments}
We would like to thank A. Ito for useful discussion. D.Y. was supported
by Grant-in-Aid for JSPS Research Fellow and JSPS KAKENHI Grant Numbers
17J00490. J.~S. was in part supported by JSPS KAKENHI Grant Numbers
JP17H02894, JP17K18778, JP15H05895, JP17H06359, JP18H04589. J.~S
is also supported by JSPS Bilateral Joint Research Projects (JSPS-NRF
collaboration) “String Axion Cosmology.” 
\end{acknowledgments}

\section*{Appendix}

To perform 6-th order WKB-method, we need $\Omega_{3},\Omega_{4}$
and $\Omega_{5}$ in addition to $\Omega_{1},\Omega_{2}$. Here, we
dispaly them for completeness:

\begin{align}
\Omega_{3} & =\left(-463020\beta_{1}^{6}+465300\beta_{1}^{4}\beta_{2}-78120\beta_{1}^{3}\beta_{3}-99780\beta_{1}^{2}\beta_{2}^{2}+10860\beta_{1}^{2}\beta_{4}+19320\beta_{1}\beta_{2}\beta_{3}\right.\nonumber \\
 & \hspace{1em}\left.-1260\beta_{1}\beta_{5}+1500\beta_{2}^{3}-660\beta_{2}\beta_{4}-630\beta_{3}^{2}+70\beta_{6}\right)\beta^{4}\nonumber \\
\nonumber \\
 & \hspace{1em}+\left(-418110\beta_{1}^{6}+479970\beta_{1}^{4}\beta_{2}-95460\beta_{1}^{3}\beta_{3}-124026\beta_{1}^{2}\beta_{2}^{2}+17070\beta_{1}^{2}\beta_{4}+29340\beta_{1}\beta_{2}\beta_{3}\right.\nonumber \\
 & \hspace{1em}\left.+3414\beta_{2}^{3}-2730\beta_{1}\beta_{5}-1770\beta_{2}\beta_{4}-1085\beta_{3}^{2}+245\beta_{6}\right)\beta^{2}\nonumber \\
\nonumber \\
 & \hspace{1em}-\frac{101479}{4}\beta_{1}^{6}+\frac{131817}{4}\beta_{1}^{4}\beta_{2}-\frac{14777}{2}\beta_{1}^{3}\beta_{3}-\frac{40261}{4}\beta_{1}^{2}\beta_{2}^{2}+\frac{6055}{4}\beta_{1}^{2}\beta_{4}+\frac{5667}{2}\beta_{1}\beta_{2}\beta_{3}\nonumber \\
 & \hspace{1em}-\frac{1155}{4}\beta_{1}\beta_{5}+\frac{1539}{4}\beta_{2}^{3}-\frac{945}{4}\beta_{2}\beta_{4}-\frac{1107}{8}\beta_{3}^{2}+\frac{315}{8}\beta_{6},
\end{align}
\begin{align}
\Omega_{4} & =\left(-95872644\beta_{1}^{8}+130619664\beta_{1}^{6}\beta_{2}-22467312\beta_{1}^{5}\beta_{3}-51067800\beta_{1}^{4}\beta_{2}^{2}+3454920\beta_{1}^{4}\beta_{4}\right.\nonumber \\
 & \hspace{1em}+13073760\beta_{1}^{3}\beta_{2}\beta_{3}+5418000\beta_{1}^{2}\beta_{2}^{3}-493920\beta_{1}^{3}\beta_{5}-1285200\beta_{1}^{2}\beta_{2}\beta_{4}-732480\beta_{1}^{2}\beta_{3}^{2}\nonumber \\
 & \hspace{1em}-1140720\beta_{1}\beta_{2}^{2}\beta_{3}-42756\beta_{2}^{4}+59472\beta_{1}^{2}\beta_{6}+98784\beta_{1}\beta_{2}\beta_{5}+110544\beta_{1}\beta_{3}\beta_{4}+25032\beta_{2}^{2}\beta_{4}\nonumber \\
 & \hspace{1em}\left.+49392\beta_{2}\beta_{3}^{2}-5544\beta_{1}\beta_{7}-3024\beta_{2}\beta_{6}-5544\beta_{3}\beta_{5}-1572\beta_{4}^{2}+252\beta_{8}\right)\beta^{5}\nonumber \\
\nonumber \\
 & \hspace{1em}+\left(-154601370\beta_{1}^{8}+231728040\beta_{1}^{6}\beta_{2}-45019560\beta_{1}^{5}\beta_{3}-101714460\beta_{1}^{4}\beta_{2}^{2}+8269260\beta_{1}^{4}\beta_{4}\right.\nonumber \\
 & \hspace{1em}+29638800\beta_{1}^{3}\beta_{2}\beta_{3}+12782760\beta_{1}^{2}\beta_{2}^{3}-1456560\beta_{1}^{3}\beta_{5}-3618360\beta_{1}^{2}\beta_{2}\beta_{4}-1870400\beta_{1}^{2}\beta_{3}^{2}\nonumber \\
 & \hspace{1em}-3138600\beta_{1}\beta_{2}^{2}\beta_{3}-178330\beta_{2}^{4}+223720\beta_{1}^{2}\beta_{6}+361200\beta_{1}\beta_{2}\beta_{5}+354120\beta_{1}\beta_{3}\beta_{4}\nonumber \\
 & \hspace{1em}\left.+118220\beta_{2}^{2}\beta_{4}+150360\beta_{2}\beta_{3}^{2}-28140\beta_{1}\beta_{7}-17640\beta_{2}\beta_{6}-21420\beta_{3}\beta_{5}-8290\beta_{4}^{2}+1890\beta_{8}\right)\beta^{3}\nonumber \\
\nonumber \\
 & \hspace{1em}+\left(-\frac{129443349}{4}\beta_{1}^{8}+53574549\beta_{1}^{6}\beta_{2}-11535783\beta_{1}^{5}\beta_{3}-\frac{53000175}{2}\beta_{1}^{4}\beta_{2}^{2}+\frac{4785249}{2}\beta_{1}^{4}\beta_{4}\right.\nonumber \\
 & \hspace{1em}+8708550\beta_{1}^{3}\beta_{2}\beta_{3}+3909285\beta_{1}^{2}\beta_{2}^{3}-482622\beta_{1}^{3}\beta_{5}-1246797\beta_{1}^{2}\beta_{2}\beta_{4}-632340\beta_{1}^{2}\beta_{3}^{2}\nonumber \\
 & \hspace{1em}-1119415\beta_{1}\beta_{2}^{2}\beta_{3}-\frac{305141}{4}\beta_{2}^{4}+88753\beta_{1}^{2}\beta_{6}+149478\beta_{1}\beta_{2}\beta_{5}+145417\beta_{1}\beta_{3}\beta_{4}+\frac{117281}{2}\beta_{2}^{2}\beta_{4}\nonumber \\
 & \hspace{1em}\left.+64731\beta_{2}\beta_{3}^{2}-\frac{28077}{2}\beta_{1}\beta_{7}-10521\beta_{2}\beta_{6}-\frac{22029}{2}\beta_{3}\beta_{5}-\frac{19277}{4}\beta_{4}^{2}+\frac{5607}{4}\beta_{8}\right)\beta\hspace{1em}\text{and}
\end{align}
\begin{align*}
\Omega_{5} & =\left(-22598568720\beta_{1}^{10}+38797354512\beta_{1}^{8}\beta_{2}-6749494080\beta_{1}^{7}\beta_{3}-22002812832\beta_{1}^{6}\beta_{2}^{2}\right.\\
 & \hspace{1em}+1080123744\beta_{1}^{6}\beta_{4}+6257684160\beta_{1}^{5}\beta_{2}\beta_{3}+4660027680\beta_{1}^{4}\beta_{2}^{3}-165561984\beta_{1}^{5}\beta_{5}\\
 & \hspace{1em}-766795680\beta_{1}^{4}\beta_{2}\beta_{4}-413669760\beta_{1}^{4}\beta_{3}^{2}-1443234240\beta_{1}^{3}\beta_{2}^{2}\beta_{3}-291804240\beta_{1}^{2}\beta_{2}^{4}\\
 & \hspace{1em}+23163840\beta_{1}^{4}\beta_{6}+85128960\beta_{1}^{3}\beta_{2}\beta_{5}+90350400\beta_{1}^{3}\beta_{3}\beta_{4}+108679200\beta_{1}^{2}\beta_{2}^{2}\beta_{4}\\
 & \hspace{1em}+126329280\beta_{1}^{2}\beta_{2}\beta_{3}^{2}+64196160\beta_{1}\beta_{2}^{3}\beta_{3}+1400784\beta_{2}^{5}-2919840\beta_{1}^{3}\beta_{7}-7531776\beta_{1}^{2}\beta_{2}\beta_{6}\\
 & \hspace{1em}-8618400\beta_{1}^{2}\beta_{3}\beta_{5}-3939600\beta_{1}^{2}\beta_{4}^{2}-6233472\beta_{1}\beta_{2}^{2}\beta_{5}-13981632\beta_{1}\beta_{2}\beta_{3}\beta_{4}-2849280\beta_{1}\beta_{3}^{3}\\
 & \hspace{1em}-1023456\beta_{2}^{3}\beta_{4}-3116736\beta_{2}^{2}\beta_{3}^{2}+307440\beta_{1}^{2}\beta_{8}+487872\beta_{1}\beta_{2}\beta_{7}+583520\beta_{1}\beta_{3}\beta_{6}\\
 & \hspace{1em}+544320\beta_{1}\beta_{4}\beta_{5}+129472\beta_{2}^{2}\beta_{6}+487872\beta_{2}\beta_{3}\beta_{5}+134736\beta_{2}\beta_{4}^{2}+272160\beta_{3}^{2}\beta_{4}\\
 & \hspace{1em}\left.-24024\beta_{1}\beta_{9}-13440\beta_{2}\beta_{8}-24024\beta_{3}\beta_{7}-14224\beta_{4}\beta_{6}-12012\beta_{5}^{2}+924\beta_{10}\right)\beta^{6}\\
\\
 & \hspace{1em}+\left(-57626387280\beta_{1}^{10}+106553134800\beta_{1}^{8}\beta_{2}-20386144800\beta_{1}^{7}\beta_{3}-65772661920\beta_{1}^{6}\beta_{2}^{2}\right.\\
 & \hspace{1em}+3750215280\beta_{1}^{6}\beta_{4}+20631693600\beta_{1}^{5}\beta_{2}\beta_{3}+15479738400\beta_{1}^{4}\beta_{2}^{3}-675647280\beta_{1}^{5}\beta_{5}\\
 & \hspace{1em}-2947719600\beta_{1}^{4}\beta_{2}\beta_{4}-1496632200\beta_{1}^{4}\beta_{3}^{2}-5324508000\beta_{1}^{3}\beta_{2}^{2}\beta_{3}-1135963920\beta_{1}^{2}\beta_{2}^{4}\\
 & \hspace{1em}+113720040\beta_{1}^{4}\beta_{6}+390240480\beta_{1}^{3}\beta_{2}\beta_{5}+381413280\beta_{1}^{3}\beta_{3}\beta_{4}+487029840\beta_{1}^{2}\beta_{2}^{2}\beta_{4}\\
 & \hspace{1em}+513631440\beta_{1}^{2}\beta_{2}\beta_{3}^{2}+283029600\beta_{1}\beta_{2}^{3}\beta_{3}+9396240\beta_{2}^{5}-17603880\beta_{1}^{3}\beta_{7}-42912240\beta_{1}^{2}\beta_{2}\beta_{6}\\
 & \hspace{1em}-43022280\beta_{1}^{2}\beta_{3}\beta_{5}-19968240\beta_{1}^{2}\beta_{4}^{2}-34807920\beta_{1}\beta_{2}^{2}\beta_{5}-69789120\beta_{1}\beta_{2}\beta_{3}\beta_{4}\\
 & \hspace{1em}-12597200\beta_{1}\beta_{3}^{3}-7583760\beta_{2}^{3}\beta_{4}-14953960\beta_{2}^{2}\beta_{3}^{2}+2324700\beta_{1}^{2}\beta_{8}+3618720\beta_{1}\beta_{2}\beta_{7}\\
 & \hspace{1em}+3614520\beta_{1}\beta_{3}\beta_{6}+3353280\beta_{1}\beta_{4}\beta_{5}+1142120\beta_{2}^{2}\beta_{6}+2859360\beta_{2}\beta_{3}\beta_{5}+1092720\beta_{2}\beta_{4}^{2}\\
 & \hspace{1em}+1464400\beta_{3}^{2}\beta_{4}-237930\beta_{1}\beta_{9}-147000\beta_{2}\beta_{8}-182490\beta_{3}\beta_{7}-135380\beta_{4}\beta_{6}\\
 & \hspace{1em}+\left.-82005\beta_{5}^{2}+12705\beta_{10}\right)\beta^{4}
\end{align*}
\begin{align}
 & \hspace{1em}+\left(-26541790065\beta_{1}^{10}+53237904993\beta_{1}^{8}\beta_{2}-11123381220\beta_{1}^{7}\beta_{3}-36045764154\beta_{1}^{6}\beta_{2}^{2}\right.\nonumber \\
 & \hspace{1em}+2279955006\beta_{1}^{6}\beta_{4}+12440307420\beta_{1}^{5}\beta_{2}\beta_{3}+9481289682\beta_{1}^{4}\beta_{2}^{3}-461383776\beta_{1}^{5}\beta_{5}\nonumber \\
 & \hspace{1em}-2012614434\beta_{1}^{4}\beta_{2}\beta_{4}-999867660\beta_{1}^{4}\beta_{3}^{2}-3630132780\beta_{1}^{3}\beta_{2}^{2}\beta_{3}-809619141\beta_{1}^{2}\beta_{2}^{4}\nonumber \\
 & \hspace{1em}+89013120\beta_{1}^{4}\beta_{6}+304548384\beta_{1}^{3}\beta_{2}\beta_{5}+292426020\beta_{1}^{3}\beta_{3}\beta_{4}+388974714\beta_{1}^{2}\beta_{2}^{2}\beta_{4}\nonumber \\
 & \hspace{1em}+392853060\beta_{1}^{2}\beta_{2}\beta_{3}^{2}+230221620\beta_{1}\beta_{2}^{3}\beta_{3}+9317949\beta_{2}^{5}-16119726\beta_{1}^{3}\beta_{7}\nonumber \\
 & \hspace{1em}-39660264\beta_{1}^{2}\beta_{2}\beta_{6}-38201310\beta_{1}^{2}\beta_{3}\beta_{5}-18168321\beta_{1}^{2}\beta_{4}^{2}-33071472\beta_{1}\beta_{2}^{2}\beta_{5}\nonumber \\
 & \hspace{1em}-63944892\beta_{1}\beta_{2}\beta_{3}\beta_{4}-10841880\beta_{1}\beta_{3}^{3}-8518614\beta_{2}^{3}\beta_{4}-14034096\beta_{2}^{2}\beta_{3}^{2}+2582685\beta_{1}^{2}\beta_{8}\nonumber \\
 & \hspace{1em}+4096764\beta_{1}\beta_{2}\beta_{7}+3870930\beta_{1}\beta_{3}\beta_{6}+3671892\beta_{1}\beta_{4}\beta_{5}+1518048\beta_{2}^{2}\beta_{6}+3168732\beta_{2}\beta_{3}\beta_{5}\nonumber \\
 & \hspace{1em}+1390869\beta_{2}\beta_{4}^{2}+1530210\beta_{3}^{2}\beta_{4}-\frac{671517}{2}\beta_{1}\beta_{9}-236460\beta_{2}\beta_{8}-\frac{476973}{2}\beta_{3}\beta_{7}\nonumber \\
 & \hspace{1em}+\left.-204771\beta_{4}\beta_{6}-\frac{444381}{4}\beta_{5}^{2}+\frac{101409}{4}\beta_{10}\right)\beta^{2}\nonumber \\
\nonumber \\
 & \hspace{1em}-\frac{2375536317}{2}\beta_{1}^{10}+\frac{5112354429}{2}\beta_{1}^{8}\beta_{2}-570170440\beta_{1}^{7}\beta_{3}-1875235809\beta_{1}^{6}\beta_{2}^{2}\nonumber \\
 & \hspace{1em}+125451228\beta_{1}^{6}\beta_{4}+697320300\beta_{1}^{5}\beta_{2}\beta_{3}+542138237\beta_{1}^{4}\beta_{2}^{3}-27429003\beta_{1}^{5}\beta_{5}\nonumber \\
 & \hspace{1em}-122723430\beta_{1}^{4}\beta_{2}\beta_{4}-\frac{121918445}{2}\beta_{1}^{4}\beta_{3}^{2}-226646440\beta_{1}^{3}\beta_{2}^{2}\beta_{3}-\frac{104283313}{2}\beta_{1}^{2}\beta_{2}^{4}\nonumber \\
 & \hspace{1em}+\frac{11623829}{2}\beta_{1}^{4}\beta_{6}+20366894\beta_{1}^{3}\beta_{2}\beta_{5}+19607424\beta_{1}^{3}\beta_{3}\beta_{4}+27070372\beta_{1}^{2}\beta_{2}^{2}\beta_{4}\nonumber \\
 & \hspace{1em}+27194427\beta_{1}^{2}\beta_{2}\beta_{3}^{2}+16533060\beta_{1}\beta_{2}^{3}\beta_{3}+\frac{1456569}{2}\beta_{2}^{5}-\frac{2336663}{2}\beta_{1}^{3}\beta_{7}-2995587\beta_{1}^{2}\beta_{2}\beta_{6}\nonumber \\
 & \hspace{1em}-\frac{5703723}{2}\beta_{1}^{2}\beta_{3}\beta_{5}-\frac{2729425\beta_{1}^{2}\beta_{4}^{2}}{2}-2594391\beta_{1}\beta_{2}^{2}\beta_{5}-5045766\beta_{1}\beta_{2}\beta_{3}\beta_{4}-854685\beta_{1}\beta_{3}^{3}\nonumber \\
 & \hspace{1em}-735210\beta_{2}^{3}\beta_{4}-\frac{2301381}{2}\beta_{2}^{2}\beta_{3}^{2}+\frac{848925}{4}\beta_{1}^{2}\beta_{8}+358344\beta_{1}\beta_{2}\beta_{7}+\frac{674037}{2}\beta_{1}\beta_{3}\beta_{6}\nonumber \\
 & \hspace{1em}+315150\beta_{1}\beta_{4}\beta_{5}+\frac{292005}{2}\beta_{2}^{2}\beta_{6}+289908\beta_{2}\beta_{3}\beta_{5}+\frac{269325}{2}\beta_{2}\beta_{4}^{2}+143370\beta_{3}^{2}\beta_{4}\nonumber \\
 & -\frac{259875}{8}\beta_{1}\beta_{9}-\frac{51975}{2}\beta_{2}\beta_{8}-\frac{203931}{8}\beta_{3}\beta_{7}-\frac{89775}{4}\beta_{4}\beta_{6}-\frac{180675}{16}\beta_{5}^{2}+\frac{51975}{16}\beta_{10}.
\end{align}

\bibliographystyle{apsrev4-1}
\bibliography{TFormField}

\end{document}